\documentclass{jfm}
\usepackage{bm}% bold math
\usepackage{graphicx}
\usepackage{epstopdf, epsfig}
\usepackage{subfigure}
\usepackage{graphics}
\usepackage{amssymb}
\usepackage{amsmath}
\usepackage{commath}
\usepackage{mathtools}
\usepackage{array}
\usepackage{color}
\usepackage{booktabs}
\usepackage[title]{appendix}

\shorttitle{The optimal displacement of immiscible two-phase fluids in a pore doublet}
\shortauthor{F. Shan, Z. H. Chai, B. C. Shi and M. Zhao}

\title{The optimal displacement of immiscible two-phase fluids in a pore doublet}

\author{Fang Shan\aff{1},
  Zhenhua Chai\aff{1,}\aff{2,}\aff{3},
  \corresp{\email{hustczh@hust.edu.cn}}
  Baochang Shi\aff{1,}\aff{2,}\aff{3}
  and Meng Zhao\aff{1,}\aff{2,}\aff{3}
 }
\affiliation
{\aff{1}School of Mathematics and Statistics, Huazhong University of Science and Technology, \\
Wuhan 430074, China \\
\aff{2}Institute of Interdisciplinary Research for Mathematics and Applied Science, \\
Huazhong University of Science and Technology, Wuhan 430074, China
\aff{3}Hubei Key Laboratory of Engineering Modeling and Scientific Computing, \\
Huazhong University of Science and Technology, Wuhan 430074, China
}

\begin{document}

\maketitle

\begin{abstract}
The displacement of multiphase fluid flow in a pore doublet is a fundamental problem, and is also of importance
in understanding of the transport mechanisms of multiphase flows in the porous media. During the displacement of
immiscible two-phase fluids in the pore doublet, the transport process is not only influenced by the capillary
and viscous forces, but also affected by the channel geometry. In this paper, we first present a mathematical
model to describe the two-phase fluid displacement in the pore doublet where the effects of capillary force,
viscous force and the geometric structure are included. Then we derive an analytical solution of the model for
the first time, and find that the displacement process is dominated by the capillary number, the viscosity ratio
and the radius ratio. Furthermore, we define the optimal displacement that the wetting fluids in two daughter
channels break through the branches simultaneously (both of them have the same breakthrough time), and also obtain
the critical capillary number corresponding to the optimal displacement, which is related to
the radius ratio of two daughter channels and viscosity ratio of two immiscible fluids. Finally, it is worthy
noting that the present analytical results on the displacement in the pore doublet can be used to explain and
understand the phenomenon of preferential imbibition or preferential flow in porous media.

\end{abstract}


\section{Introduction}\label{Introduction}
The displacement of immiscible fluids in the porous media is ubiquitous in both nature and engineering, such as
the water transport in capillaries of plants, cleaning
up the spilt liquid with a towel, oil recovery, to name but a few \citep{de2004capillarity,xu2014effect}.
However, due to the presence of the complex fluid-fluid interface and pore structure, it is not only difficult
to accurately predict displacement process, but also challenging to clearly understand the physical mechanisms
inherent in the immiscible displacement in porous media.

The pore doublet is composed of a pair of parallel daughter channels, which connect from entrance to the downstream.
As a basic pore-scale prototype of multiphase flows in porous media, the pore doublet model can be used to explain
the entrapment and distribution of the displaced and/or displacing fluid in the network of pore space, and also
receives increasing attention in the study of complex displacement process in porous media \citep{Sun2016}.
Owing to its wide applications, the pore doublet model has also been studied extensively from experimental,
theoretical and numerical perspectives. For instance, \cite{Rose1956} conducted a simple analysis on the
displacement of oil by water where the fluid viscosities are assumed to be same as each other, and demonstrated
that the oil tends to be trapped in the smaller channel of the pore doublet. \cite{Laidlaw1983} performed a
theoretical and experimental study on the mechanism for the trapping oil in a pore doublet under different
fluid viscosities, and found that the characteristics of the source and sink play a crucial role in determining
the occurrence for trapping in the wide or narrow branch of the doublet. \cite{Chatzis1983} obtained an explicit
expression of velocity under the assumption of both fluids with the same viscosity, and carried out some
experiments for the motions of capillary interfaces in the drainage-type and imbibition-type displacements.
Inspired by the work of \cite{Chatzis1983}, \cite{Sorbie1995} extended the Lucas-Washburn
equation \citep{Lucas,Washburn}, and derived a mathematical model for the oil-water
displacement in the pore doublet. Moreover, other researchers also considered the movement of displacement
interface in a pore doublet system. For example, \cite{Bokserman1991} developed a mathematical model to
analyze the movement of displacement front, and found that the viscosity ratio of displacing and displaced fluids
has an important influence on the displacement process. In addition, \cite{Lundstrom2007} studied the condition of
an overflow in a pore doublet, and illustrated that the leading front of the liquid can be in the wide or narrow
capillary channel. Recently, \cite{Al-Housseiny2014} considered the two-phase fluid displacement in a network
composed of two identical channels, and the results show that the dynamics of fluid penetration are governed
by the viscosity ratio of the fluids, the channel geometry and a critical parameter related the capillary number.

On the other hand, with the rapid development of computational fluid dynamics (CFD), the numerical simulation
has become an efficient and powerful tool in exploring the mechanism of the displacement phenomena in the pore
doublet system. \cite{Wolf2008} used Boolean lattice-gas model to simulate fluid-fluid interface displacement
inside a pore doublet, and found that the immiscible displacement is strongly affected by the contact angle,
and a smaller contact angle results in a better displacement. \cite{Jiaotao2018} proposed an improved
Shan-Chen lattice Boltzmann method (LBM) to simulate the spontaneous imbibition, and investigated the influence
of channel geometry on the displacement in the pore doublet. Recently, \cite{Gu2021} adopted the colour-gradient
LBM to study the displacement of two immiscible fluids in the pore doublet, and found that there are three
different displacement patterns, and the displacement process is influenced by the competition between the
viscous and capillary forces. \cite{Fanli2022} also used colour-gradient LBM to produce a complete phase
diagram for the preferential flow in a pore doublet, which can be extended to the dual permeable media.
Additionally, \cite{Nabizadeh2019} applied a traditional CFD approach to study the two-phase immiscible
displacement in a pore doublet, and mainly focused on effects of injection velocity, the presence of wetting
film on the walls, wettability and capillary number. The numerical results show that the presence of wetting
film and the increase of the injection velocity have a positive effect on the displacement process.

From above literature review, it is clear that the displacement of immiscible two-phase fluids in the pore doublet
is a classic problem, and has been studied extensively. To our knowledge, however, there is no analytical solution
available to this problem under different fluid viscosities and channel radii. To fill the gap, in this work,
we first present a mathematical model for the displacement of immiscible two-phase fluids in the pore doublet,
and then obtain the analytical solution of the model where the effects of the capillary force, viscous force
and the geometric structure of pore doublet are included. To depict the displacement process quantitatively,
a breakthrough time when the wetting fluids in narrow or wide channel first reaches the end of branches is studied.
Additionally, the optimal displacement is also introduced to represent the case where the
displaced fluids are completely expelled from two daughter channels with the shortest time. Furthermore,
we consider the effects of capillary number, viscosity ratio and radius ratio on the displacement pattern
and breakthrough behavior. The displacement process in the pore doublet depends on the relative relationship
between the capillary and viscous forces, and the optimal displacement occurs at a critical capillary number.
The rest of the paper is organized as follows. In section \ref{Mathematical model}, a mathematical model is
presented to describe immiscible displacement in a pore doublet. In section \ref{2D_solution}, the analytical
solution of the model and the critical capillary number for the optimal displacement are obtained, and finally,
some conclusions are summarized in section \ref{conclusion}.

\section{Mathematical model for immiscible displacement in a pore doublet}\label{Mathematical model}
\begin{figure}
\centering
\includegraphics[width=0.7\textwidth]{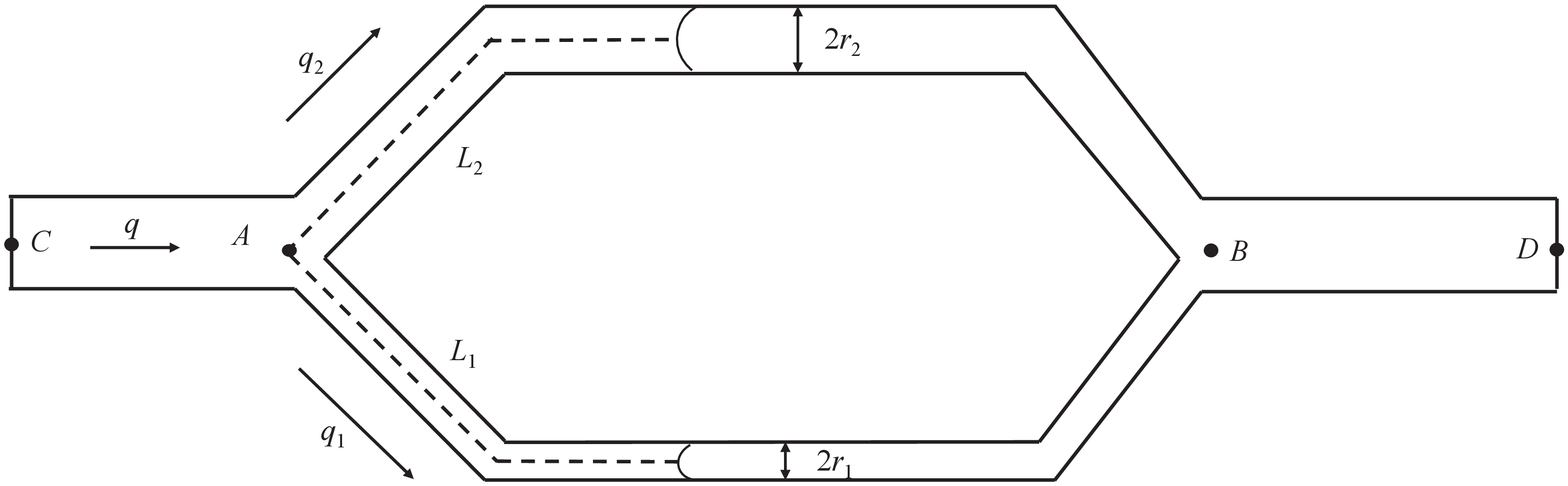}
\caption{Schematic diagram of the displacement process in a two-dimensional pore doublet.}
\label{schematic}
\end{figure}
For the displacement of immiscible two-phase fluids in the pore doublet (see Fig. \ref{schematic}), the wetting
fluid is supplied along a straight channel from the point $C$ to point $A$ with a volumetric flow rate $q$. It
flows though two capillary channels which bifurcate from point $A$ and reunite at downstream (point $B$) and finally
the fluid flows out of the straight channel $BD$. The radii of bottom and top branches (labeled by channel $1$
and channel $2$) are $r_{1}$ and $r_{2}$, respectively. The total length of each channel along flowing direction
is $L$ and the width of the straight channel $CA$ or $BD$ is $h=2(r_{1}+r_{2})$. Initially, the pore doublet is filled with
non-wetting fluid.

In the pore doublet, a laminar Poiseuille flow is assumed to be satisfied, and the two-phase interface advances with a constant mean curvature. The pressure drop $\Delta p$ across capillary branches is composed of the viscous pressure drop $\Delta p_{i}$ and capillary pressure $\Delta p_{c_{i}}$, and for two-dimensional case (see Appendix \ref{3D_model} for three-dimensional case), it can be written as \citep{Gu2021}
\begin{equation}
\Delta p=p_{A}-p_{B}=\Delta p_{i}+\Delta p_{c_{i}}=\frac{3q_{i}}{2r_{i}^3}\left[\mu_{w}L_{i}+\mu_{n}(L-L_{i})\right]-\frac{\sigma \cos\theta}{r_{i}},
\label{eq:pressure}
\end{equation}
where $p_{A}$ and $p_{B}$ are pressures at points $A$ and $B$, $q_{i}=2r_{i}\cdot u_{i}$ is the flow rate through the channel $i$, $u_{i}$ is the average velocity and $\theta$ is the contact angle. $\mu_{w}$ and $\mu_{n}$ denote the dynamic viscosities of wetting and non-wetting fluids. $\sigma$ is the interfacial tension coefficient, and $L_{i}$ is the length of the wetting fluid advancing along the capillary channel $i$.

Based on the mass conservation, the total volumetric flow rate $q$ can be given by
\begin{equation}
q=q_{1}+q_{2}.
\label{eq:mass_conserve}
\end{equation}
To simplify the following analysis, some scaling parameters are introduced,
\begin{equation}
L_{c}=r_{1},\quad T_{c}=\frac{2r_{1}L}{q},\quad P_{c}=\frac{3\mu_{n}qL}{2r_{1}^3},\quad U_{c}=\frac{L_{c}}{T_{c}},
\label{eq:parameters}
\end{equation}
which are used to determine the following dimensionless variables,
\begin{equation}
\hat{r}_{i}=\frac{r_{i}}{L_{c}},\quad \hat{L}_{i}=\frac{L_{i}}{L_{c}},\quad \hat{L}=\frac{L}{L_{c}},\quad \hat{t}=\frac{t}{T_c},\quad \hat{u}_{i}=\frac{u_{i}}{U_{c}},\quad \hat{q}_{i}=\frac{q_{i}}{q}.
\label{eq:variable}
\end{equation}
With the help of (\ref{eq:variable}), we obtain the dimensionless form of pressure drop from (\ref{eq:pressure}),
\begin{equation}
\begin{aligned}
\Delta \hat{p}&=\Delta \hat{p}_{i}+\Delta \hat{p}_{c_{i}}\\
&=\hat{q}_{i}\left[\frac{\eta\cdot\left(\frac{\hat{L}_{i}}{\hat{L}}\right)}{\hat{r}_{i}^3}+\frac{1-\frac{\hat{L}_{i}}{\hat{L}}}{\hat{r}_{i}^3}\right]-
\frac{1}{Ca}\frac{1}{\hat{r}_{i}},\\
&=\hat{r}_{i}\cdot\frac{d\left(\frac{\hat{L}_{i}}{\hat{L}}\right)}{d\hat{t}}
\left[\eta\frac{\left(\frac{\hat{L}_{i}}{\hat{L}}\right)}{\hat{r}_{i}^3}+\frac{\left(1-\frac{\hat{L}_{i}}{\hat{L}}\right)}{\hat{r}_{i}^3}\right]-\frac{1}{Ca}\cdot\frac{1}{\hat{r}_{i}},\quad i=1,2,\\
\end{aligned}
\label{eq:dp}
\end{equation}
where $\eta=\mu_{w}/\mu_{n}$ is the viscosity ratio, $Ca=3\mu_{n}qL/(2L_{c}^2\sigma\cos\theta)$ is the capillary number and $\hat{q}_{i}=\hat{r}_{i}\hat{u}_{i}/\hat{L}=\hat{r}_{i}\frac{d\hat{L}_{i}}{d\hat{t}}\frac{1}{\hat{L}}=\hat{r}_{i}d\left(\frac{\hat{L}_{i}}{\hat{L}}\right)/d\hat{t}$.

Substituting expression of $\hat{q}_{i}$ into (\ref{eq:mass_conserve}), we derive the dimensionless form of mass conserve equation as
\begin{equation}
\begin{aligned}
\frac{d\left(\frac{\hat{L}_{1}}{\hat{L}}\right)}{d\hat{t}}\cdot\hat{r}_{1}+\frac{d\left(\frac{\hat{L}_{2}}{\hat{L}}\right)}{d\hat{t}}\cdot\hat{r}_{2}=1.
\label{eq:d_mass}
\end{aligned}
\end{equation}
Based on the fact that the two branches are interconnected at both ends, the net pressure drop over either of the
two capillary channels $1$ and $2$ between points $A$ and $B$ must be the same, which leads to the following equation,
\begin{equation}
\begin{aligned}
&\hat{r}_{1}\cdot\frac{d\left(\frac{\hat{L}_{1}}{\hat{L}}\right)}{d\hat{t}}
\left[\eta\frac{\left(\frac{\hat{L}_{1}}{\hat{L}}\right)}{\hat{r}_{1}^3}+\frac{\left(1-\frac{\hat{L}_{1}}{\hat{L}}\right)}{\hat{r}_{1}^3}\right]-\frac{1}{Ca}\cdot\frac{1}{\hat{r}_{1}}\\
&=\hat{r}_{2}\cdot\frac{d\left(\frac{\hat{L}_{2}}{\hat{L}}\right)}{d\hat{t}}
\left[\eta\frac{\left(\frac{\hat{L}_{2}}{\hat{L}}\right)}{\hat{r}_{2}^3}+\frac{\left(1-\frac{\hat{L}_{2}}{\hat{L}}\right)}{\hat{r}_{2}^3}\right]-\frac{1}{Ca}\cdot\frac{1}{\hat{r}_{2}},
\end{aligned}
\label{eq:duaghter_channel}
\end{equation}
where $\hat{r}_{1} = 1$ and the radius ratio $\hat{r}_{2}=r_{2}/r_{1}$ is denoted by $\hat{r}$ in the following analysis. Noted
that (\ref{eq:d_mass}) and (\ref{eq:duaghter_channel}) are governing equations for the displacement in
the two-dimensional pore doublet.
\section{The analytical solution of mathematical model and the critical capillary number for the optimal displacement}\label{2D_solution}
In this section, we first present a detailed process to obtain the analytical solution of the governing equations (\ref{eq:d_mass}) and (\ref{eq:duaghter_channel}), and then determine the critical capillary number for the optimal displacement in the pore doublet.
\subsection{The analytical solution of mathematical model}
According to (\ref{eq:d_mass}) and (\ref{eq:duaghter_channel}), one can obtain the following dimensionless ordinary differential equations in terms of $\hat{L}_{i}(t)\ (i=1,2)$,
\begin{subequations}
\begin{equation}
\frac{d\left(\frac{\hat{L}_{1}}{\hat{L}}\right)}{d\hat{t}}=\frac{\frac{1}{Ca}\cdot\left(1-\frac{1}{\hat{r}}\right)+\phi \left(\frac{\hat{L}_{2}}{\hat{L}}\right)}{\left[\phi\left(\frac{\hat{L}_{1}}{\hat{L}}\right)+\phi\left(\frac{\hat{L}_{2}}{\hat{L}}\right)\right]}, \label{eq:1a}
\end{equation}
\begin{equation}
\frac{d\left(\frac{\hat{L}_{2}}{\hat{L}}\right)}{d\hat{t}}=\frac{-\frac{1}{Ca}\cdot\left(1-\frac{1}{\hat{r}}\right)+\phi \left(\frac{\hat{L}_{1}}{\hat{L}}\right)}{\hat{r}\left[\phi\left(\frac{\hat{L}_{1}}{\hat{L}}\right)+\phi\left(\frac{\hat{L}_{2}}{\hat{L}}\right)\right]},\label{eq:1b}
\end{equation}
\label{eq:1}
\end{subequations}
where $\phi\left(\hat{L}_{i}/\hat{L}\right)=\left[\eta\hat{L}_{i}/\hat{L}+\left(1-\hat{L}_{i}/\hat{L}\right)\right]/\hat{r}_{i}^3$. For brevity, we introduce the variable $x_{i}(\hat{t})=\hat{L}_{i}/\hat{L}$ and the parameter $Ca_{m}=\frac{1}{Ca}\cdot\left(1-\frac{1}{\hat{r}}\right)$ to rewrite (\ref{eq:1}) into the following form,
\begin{subequations}
\begin{equation}
x_{1}'(\hat{t})=\frac{Ca_{m}+\frac{\eta-1}{\hat{r}^3}x_{2}(\hat{t})+\frac{1}{\hat{r}^3}}{(\eta-1)\left[x_{1}(\hat{t})+\frac{x_{2}(\hat{t})}{\hat{r}^3}\right]+1+\frac{1}{\hat{r}^3}},\label{eq:2a}
\end{equation}
\begin{equation}
\hat{r}x_{2}'(\hat{t})=\frac{-Ca_{m}+(\eta-1)x_{1}(\hat{t})+1}{(\eta-1)\left[x_{1}(\hat{t})+\frac{x_{2}(\hat{t})}{\hat{r}^3}\right]+1+\frac{1}{\hat{r}^3}}.\label{eq:2b}
\end{equation}
\label{eq:2}
\end{subequations}
After introducing the variables $\bar{x}_{1}(\hat{t})=x_{1}(\hat{t})$ and $\bar{x}_{2}(\hat{t})= x_{2}(\hat{t})/\hat{r}^3$, (\ref{eq:2}) can be written as
\begin{subequations}
\begin{equation}
\bar{x}_{1}'(\hat{t})=\frac{Ca_{m}+(\eta-1)\bar{x}_{2}(\hat{t})+\frac{1}{\hat{r}^3}}{(\eta-1)\left[\bar{x}_{1}(\hat{t})+\bar{x}_{2}(\hat{t})\right]+1+\frac{1}{\hat{r}^3}},\label{eq:3a}
\end{equation}
\begin{equation}
\hat{r}^4\bar{x}_{2}'(\hat{t})=\frac{-Ca_{m}+(\eta-1)\bar{x}_{1}(\hat{t})+1}{(\eta-1)\left[\bar{x}_{1}(\hat{t})+\bar{x}_{2}(\hat{t})\right]+1+\frac{1}{\hat{r}^3}}.\label{eq:3b}
\end{equation}
\label{eq:3}
\end{subequations}
From (\ref{eq:3a}) and (\ref{eq:3b}), one obtains
\begin{subequations}
\begin{equation}
\bar{x}_{1}'(\hat{t})+\hat{r}^4\bar{x}_{2}'(\hat{t})=1,\label{eq:4a}
\end{equation}
\begin{equation}
(\eta-1)\bar{x}_{1}(\hat{t})\bar{x}_{1}'(\hat{t})+(1-Ca_{m})\bar{x}_{1}'(\hat{t})=\hat{r}^4(\eta-1)\bar{x}_{2}(\hat{t})\bar{x}_{2}'(\hat{t})+(\hat{r}^4Ca_{m}+\hat{r})\bar{x}_{2}'(\hat{t}).\label{eq:4b}
\end{equation}
\label{eq:4}
\end{subequations}
We integrate (\ref{eq:4}) with initial conditions $\bar{x}_{1}(0)=0$ and $\bar{x}_{2}(0)=0$, and it yields
\begin{subequations}
\begin{equation}
\bar{x}_{1}(\hat{t})=\hat{t}-\hat{r}^4\bar{x}_{2}(\hat{t}),\label{eq:5a}
\end{equation}
\begin{equation}
\frac{(\eta-1)}{2}\bar{x}_{1}(\hat{t})^2+(1-Ca_{m})\bar{x}_{1}(\hat{t})=\frac{\hat{r}^4}{2}(\eta-1)\bar{x}_{2}(\hat{t})^2+(\hat{r}^4Ca_{m}+\hat{r})\bar{x}_{2}(\hat{t}).\label{eq:5b}
\end{equation}
\label{eq:5}
\end{subequations}
Substituting (\ref{eq:5a}) into (\ref{eq:5b}), we are able to derive the following quadratic equation of $\bar{x}_{2}(\hat{t})$,
\begin{equation}
\frac{\eta-1}{2}\hat{r}^4(\hat{r}^4-1)\bar{x}_{2}^2(\hat{t})-\left[(\eta-1)\hat{r}^4\hat{t}+\hat{r}^4+\hat{r}\right]\bar{x}_{2}(\hat{t})+\frac{\eta-1}{2}\hat{t}^2+(1-Ca_{m})\hat{t}=0.\label{eq:6}
\end{equation}

Based on the formula for the extraction of square roots ($\eta\neq1$) and the condition of $0\leq\bar{x}_{i}(\hat{t})\leq1$, we have
\begin{equation}
\begin{cases}
\bar{x}_{1}(\hat{t})=\hat{t}-\hat{r}^4\bar{x}_{2}(\hat{t}), \\
\bar{x}_{2}(\hat{t})=\frac{-b-\sqrt{\Delta}}{2a},\\
\end{cases}
\label{eq:root}
\end{equation}
where
\begin{equation}
\begin{aligned}
a&=\frac{\eta-1}{2}\hat{r}^4(\hat{r}^4-1), \quad b=-\left[(\eta-1)\hat{r}^4\hat{t}+\hat{r}^4+\hat{r}\right], \quad c=\frac{\eta-1}{2}\hat{t}^2+(1-Ca_{m})\hat{t},\\
\Delta&=b^2-4ac=(\eta-1)^2\hat{r}^4\hat{t}^2+2(\eta-1)\hat{r}^4\left[\hat{r}+1+(\hat{r}^4-1)Ca_{m}\right]\hat{t}+(\hat{r}^4+\hat{r})^2.\label{eq:a_b_c}
\end{aligned}
\end{equation}

Furthermore, one can also obtain displacement lengths of wetting fluids in two branches from (\ref{eq:root}),
\begin{equation}
\begin{cases}
\hat{L}_{1}=\hat{L}\ (\hat{t} - \hat{r}^4\frac{-b-\sqrt{\Delta}}{2a}), \\
\hat{L}_{2} = \hat{r}^3\hat{L}\ \frac{-b-\sqrt{\Delta}}{2a}.\\
\end{cases}
\label{eq:L1_L2}
\end{equation}
It is interesting that when the wetting fluid in the wide or narrow tube first breaks through the channel, i.e., $\hat{L}_{1}=\hat{L}$ or $\hat{L}_{2}=\hat{L}$, the penetration time $\hat{t}_{1}$ or $\hat{t}_{2}$ can be determined from (\ref{eq:L1_L2}).

In particular, if $\eta=1$, a simple form $\hat{L}_{i}$ can be obtained from (\ref{eq:1}) or (\ref{eq:5}),
\begin{equation}
\begin{cases}
\hat{L}_{1}=\hat{L}\ \frac{Ca_{m}\hat{r}^3+1}{\hat{r}^3+1}\hat{t}, \\
\hat{L}_{2}=\hat{L}\ \frac{\hat{r}^2(1-Ca_{m})}{\hat{r}^3+1}\hat{t},\\
\end{cases}
\label{eq:lam_1}
\end{equation}
where the condition $Ca>(1-1/\hat{r})$ should hold. Then the penetration time for the first breakthrough $\hat{t}_{i}$ can be simply determined as
\begin{equation}
\begin{cases}
\hat{t}_{1} = \frac{\hat{r}^3 + 1}{\hat{r}^3Ca_{m} + 1},\\
\hat{t}_{2} = \frac{\hat{r}^3 + 1}{\hat{r}^2(1-Ca_{m})}.\\
\end{cases}
\label{eq:prefer_time_eta_1}
\end{equation}
\subsection{The critical capillary number for the optimal displacement}\label{critical_Ca}
\begin{figure}
\centering
\includegraphics[width=0.7\textwidth]{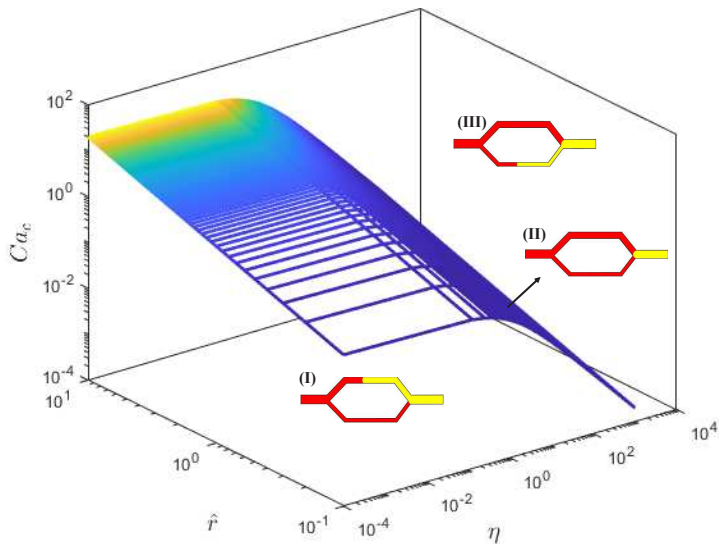}
\caption{The effects of viscosity and radius ratios on the critical capillary number.}
\label{lambda_k_Cam}
\end{figure}

We now continue to analyze the critical condition under which the wetting fluids in both capillary channels can
reach the point $B$ at the same time. Actually, when the wetting fluids in both channels simultaneously arrive at
the point $B$, the condition $x_{1}(\hat{t}) = x_{2}(\hat{t}) = 1$ or $\hat{L}_{1}=\hat{L}_{2}=\hat{L}$ is satisfied.
In this case, one derives the following critical condition from (\ref{eq:L1_L2}) and (\ref{eq:lam_1}),
\begin{equation}
Ca_{c} = \frac{2\hat{r}}{\eta + 1},
\label{eq:Cam_n_1}
\end{equation}
where $Ca_c$ is the critical capillary number corresponding to the optimal displacement, and the relation
between $Ca_{m}$ and $Ca$ has been used. It is clear from (\ref{eq:Cam_n_1}) that for a fixed viscosity ratio,
the occurrence of the critical breakthrough depends on the radius ratio $\hat{r}$ and capillary number $Ca$.
Additionally, when $\eta = 1$, the critical capillary number is equal to the radius ratio.
To see the influence of $\hat{r}$ and $\eta$ on the critical capillary number $Ca_{c}$ more clearly, we plot the
results in Fig. \ref{lambda_k_Cam} where $\hat{r}$ and $\eta$ vary from $0.1$ to $10$ and
from $10^{-4}$ to $100$, respectively. As seen from this figure and (\ref{eq:Cam_n_1}), for a given viscosity ratio $\eta$,
$Ca_c$ increases linearly in $\hat{r}$; while for a fixed radius ratio $\hat{r}$, $Ca_c$ decreases nonlinearly
with the increase of $\eta$.

\subsection{The results and discussion}
\begin{figure}
\centering
\includegraphics[width=0.5\textwidth]{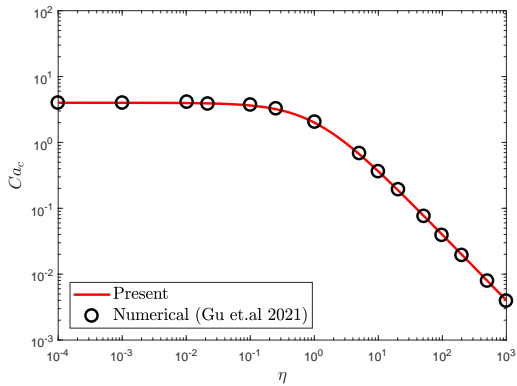}
\caption{A comparison of critical capillary number $Ca_{c}$ (\ref{eq:Cam_n_1}) with the numerical results \citep{Gu2021}. }
\label{Com_Gu}
\end{figure}

According to the critical capillary number $Ca_c$, we can predict the displacement pattern of immiscible two-phase
fluids under different values of viscosity ratio $\eta$ and radius ratio $\hat{r}$. Actually, based on the results
shown in Fig. \ref{lambda_k_Cam}, the displacement patterns of immiscible two-phase fluids in the pore doublet can
be classified into three cases: (\uppercase\expandafter{\romannumeral1}) the meniscus in narrow channel outstrips
that in wide channel at breakthrough (the regions below the surface), i.e., $\hat{L}_{1} > \hat{L}_{2}$;
(\uppercase\expandafter{\romannumeral2}) the meniscus in wide channel outstrips that in narrow at breakthrough
(the regions above the surface), i.e., $\hat{L}_{1} < \hat{L}_{2}$; and (\uppercase\expandafter{\romannumeral3})
the menisci in both channels simultaneously arrive at the downstream junction (the points on the surface),
i.e., $\hat{L}_{1} = \hat{L}_{2}$. To test the capacity of the analytical solution (\ref{eq:L1_L2}) in predicting
the displacement of immiscible two-phase fluids in the pore doublet, we carry out a comparison with the results
reported in the previous work \citep{Gu2021} where the pore-scale LBM simulations are performed. As shown in
Fig. \ref{Com_Gu}, the analytical expression of critical capillary number is in a good agreement with the
numerical results \citep{Gu2021}, which also indicates that the present analytical solution is accurate in
describing the simultaneous breakthrough of immiscible two-phase fluids in the pore doublet.

\begin{figure}
	\centering
	\includegraphics[width=0.5\textwidth]{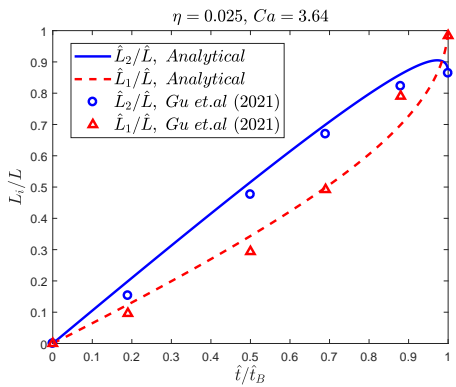}
	\caption{A comparison of penetration length between the analytical solution (\ref{eq:L1_L2}) and numerical results \citep{Gu2021}.}
	\label{Gu_net}
\end{figure}

Moreover, we also conduct a comparison between the analytical solution (\ref{eq:L1_L2}) and numerical results
of the penetration length \citep{Gu2021} in Fig. \ref{Gu_net} where $\eta=0.025$, $Ca=3.64$, $\hat{L}_{1}$
and $\hat{L}_{2}$ are normalized by $\hat{L}$, $\hat{t}$ is normalised by the dimensionless breakthrough
time $\hat{t}_{B}=t_{B}/T_{c}$, $t_{B}$ is the time when the wetting fluids first reach
point $B$, and it can be expressed by $t_{B}=\frac{2(r_{1}+r_{2})L}{q}$ when the wetting fluids simultaneously
arrive at the breakthrough point $B$. As seen from this figure, the present solution is also consistent with
the numerical data \citep{Gu2021}.

\begin{figure}
	\centering
    \subfigure[]{
		\begin{minipage}[t]{0.3\linewidth}
			\centering
			\includegraphics[width=1.8in]{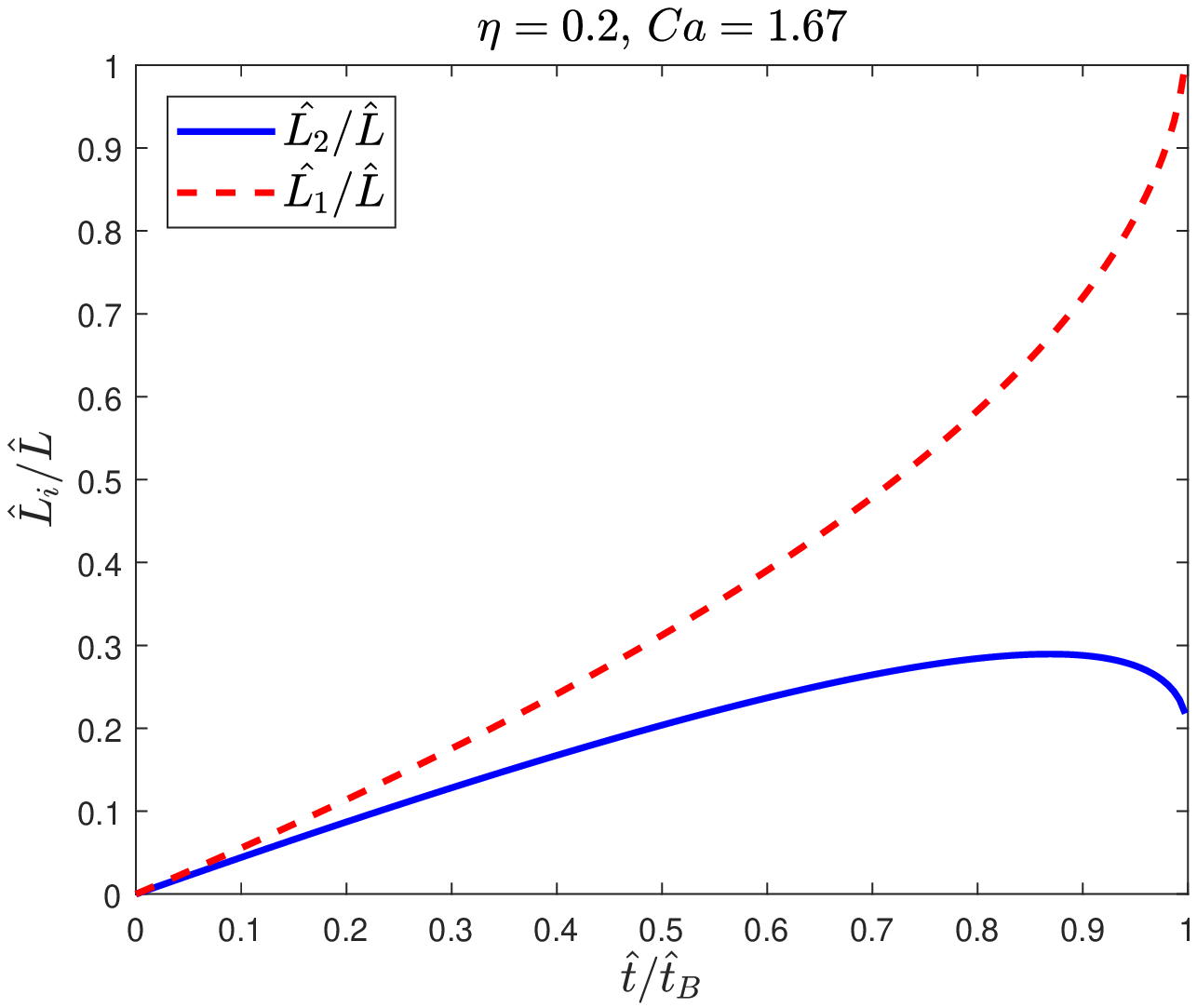}
		\end{minipage}
        \label{k_2_a}
	}
	\subfigure[]{
		\begin{minipage}[t]{0.3\linewidth}
			\includegraphics[width=1.8in]{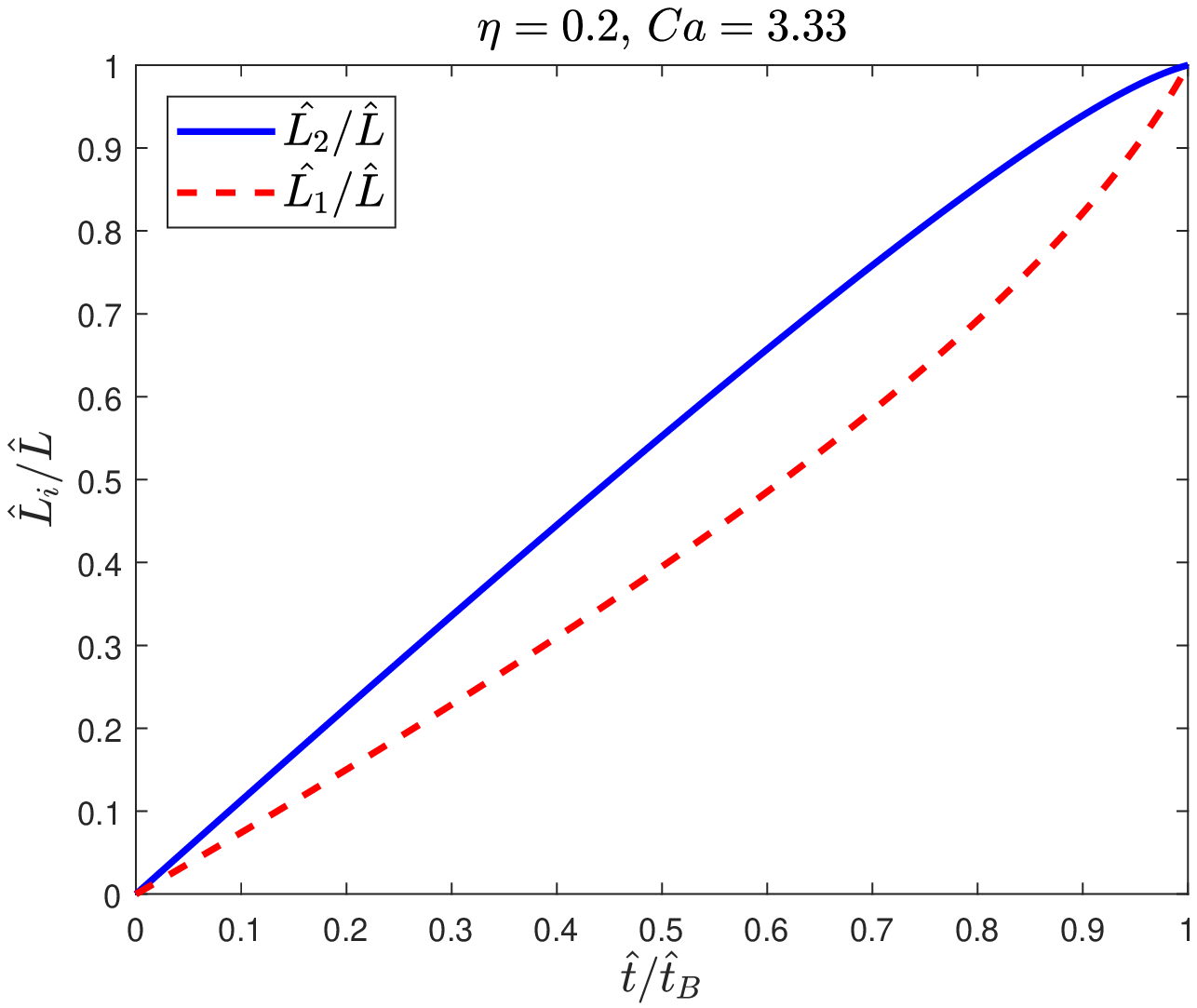}
		\end{minipage}
        \label{k_2_b}
	}
	\subfigure[]{
		\begin{minipage}[t]{0.3\linewidth}
			\centering
			\includegraphics[width=1.8in]{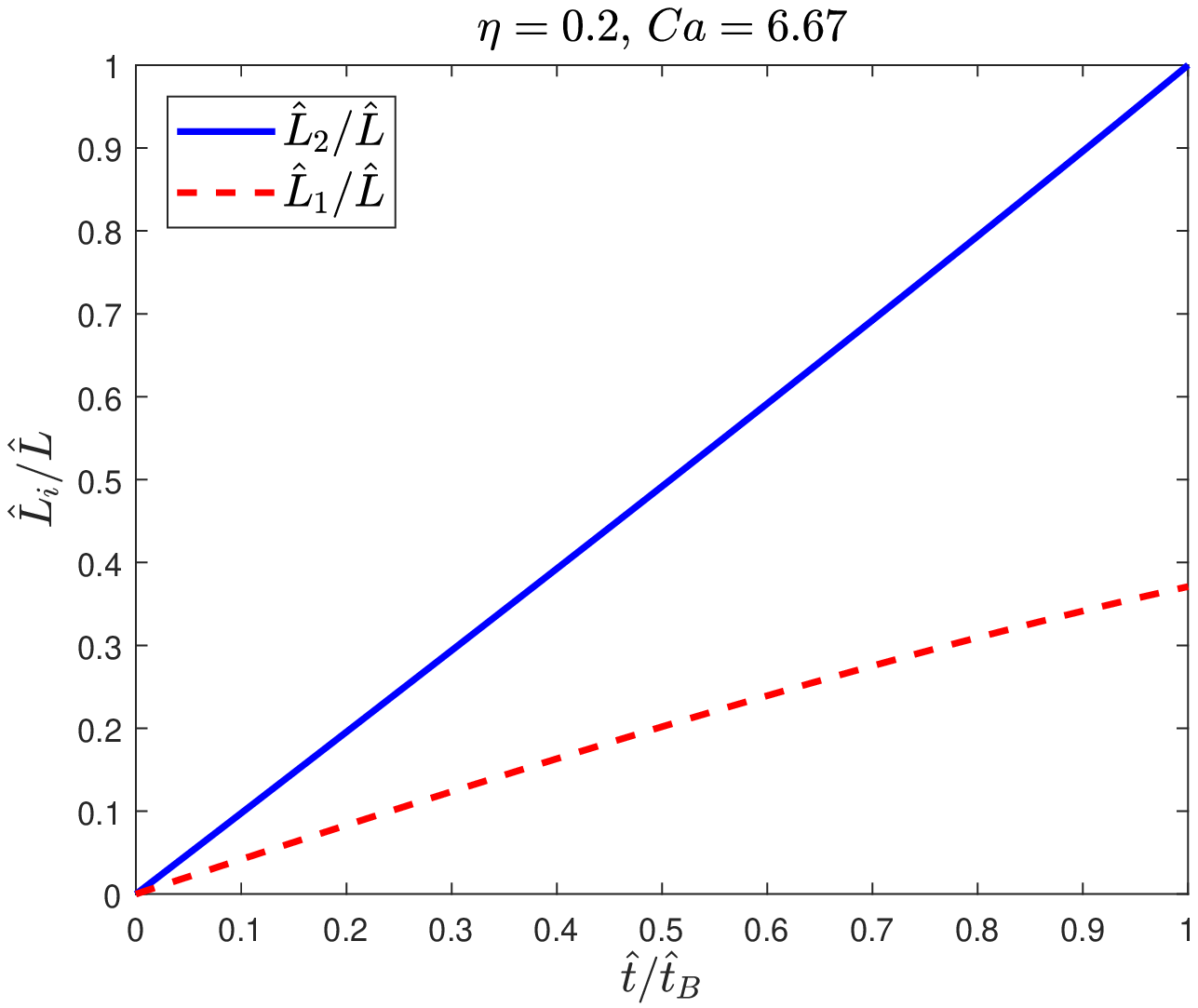}
		\end{minipage}
        \label{k_2_c}
	}

	\subfigure[]{
		\begin{minipage}[t]{0.3\linewidth}
			\centering
			\includegraphics[width=1.8in]{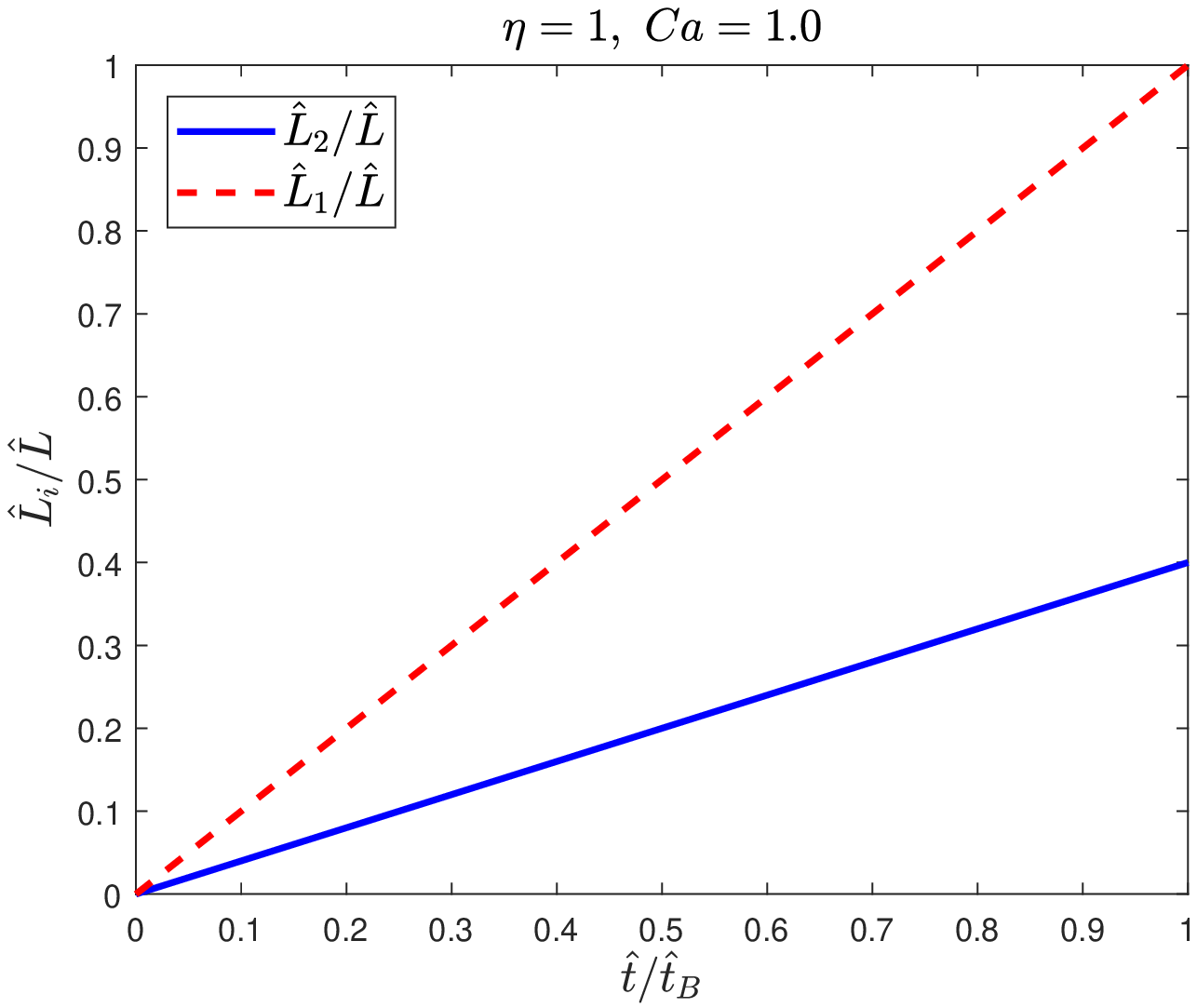}
		\end{minipage}
        \label{k_2_d}
	}
	\subfigure[]{
		\begin{minipage}[t]{0.3\linewidth}
			\includegraphics[width=1.8in]{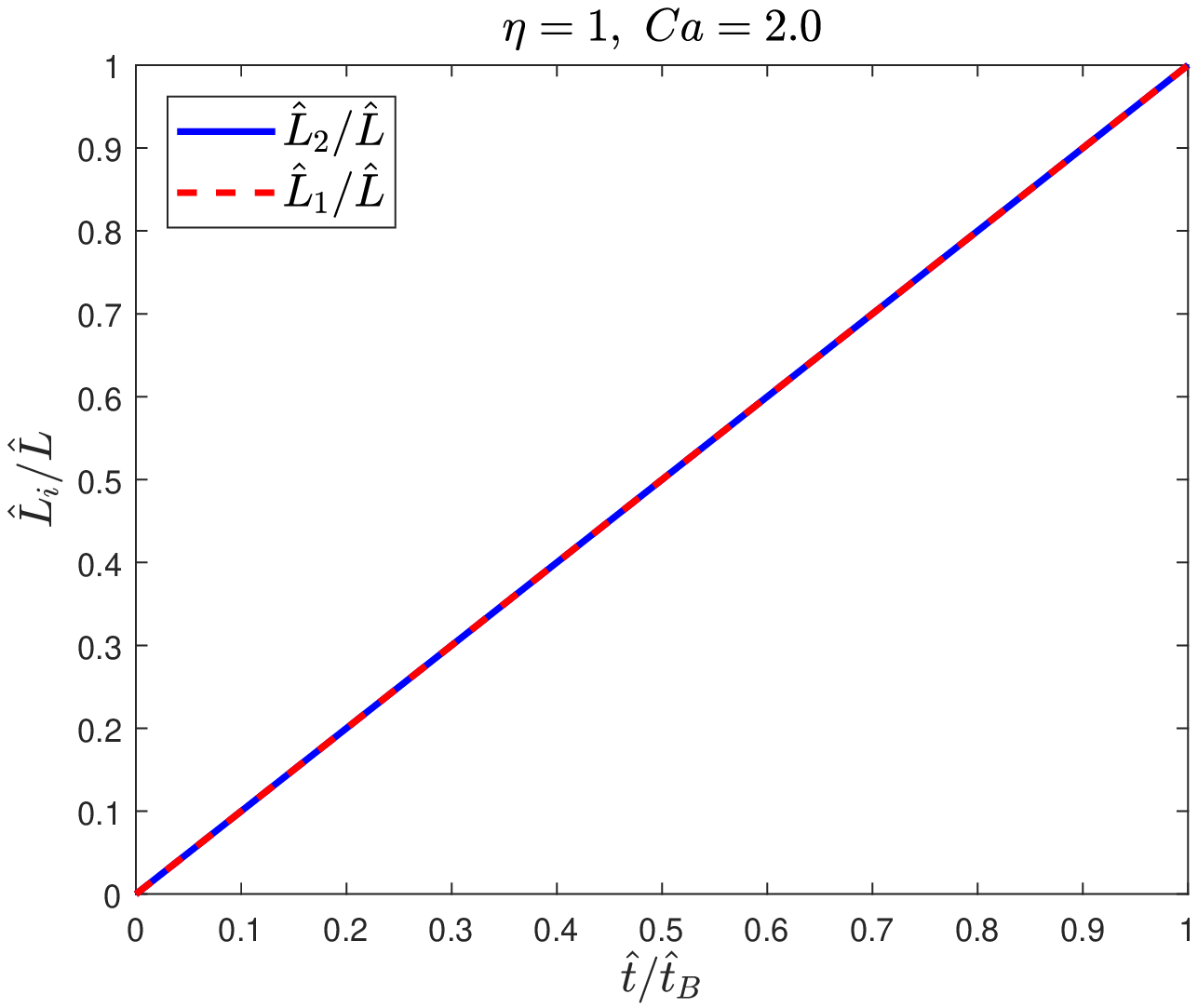}
		\end{minipage}
        \label{k_2_e}
	}
	\subfigure[]{
		\begin{minipage}[t]{0.3\linewidth}
			\centering
			\includegraphics[width=1.8in]{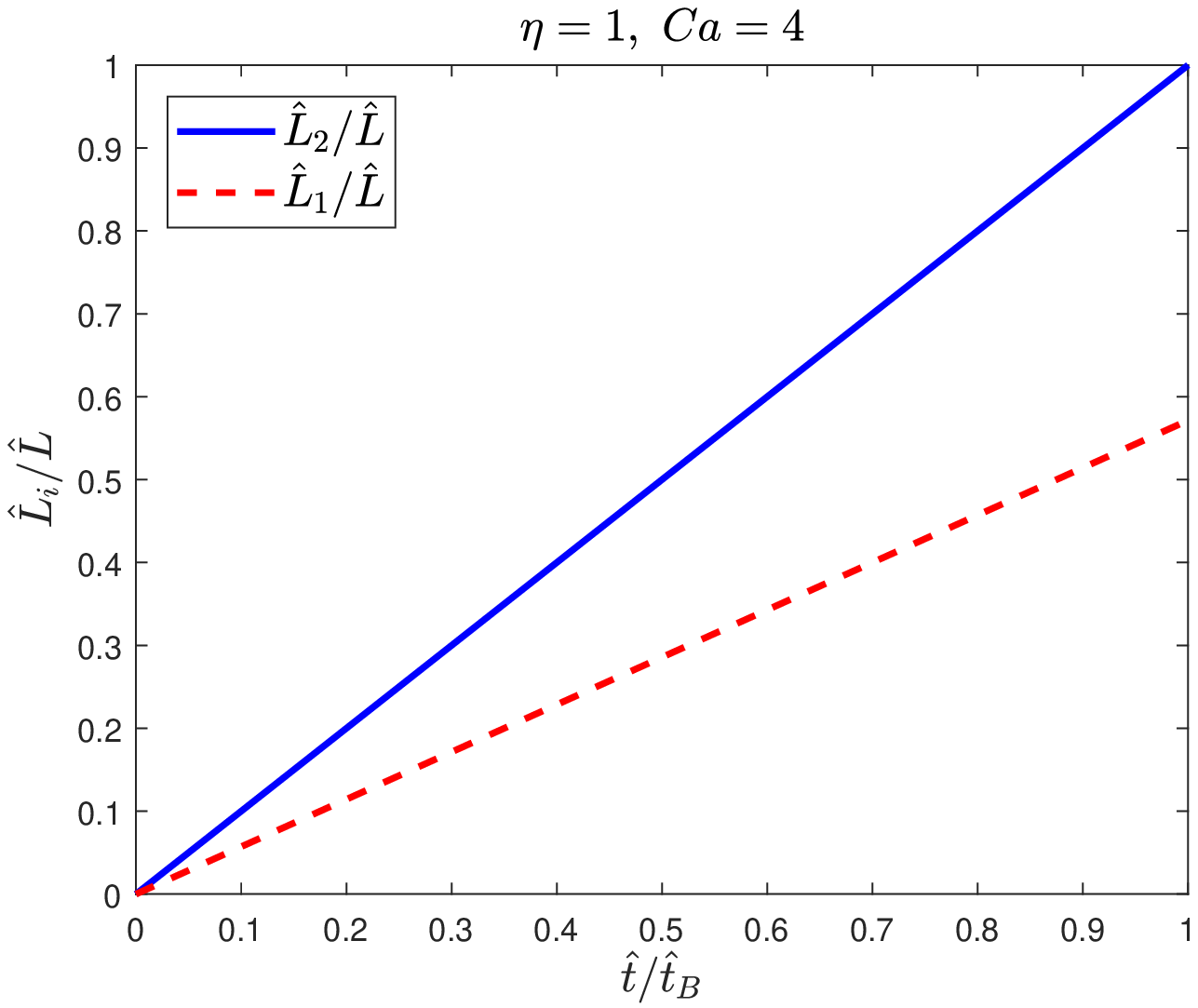}
		\end{minipage}
        \label{k_2_f}
	}

\subfigure[]{
		\begin{minipage}[t]{0.3\linewidth}
			\centering
			\includegraphics[width=1.8in]{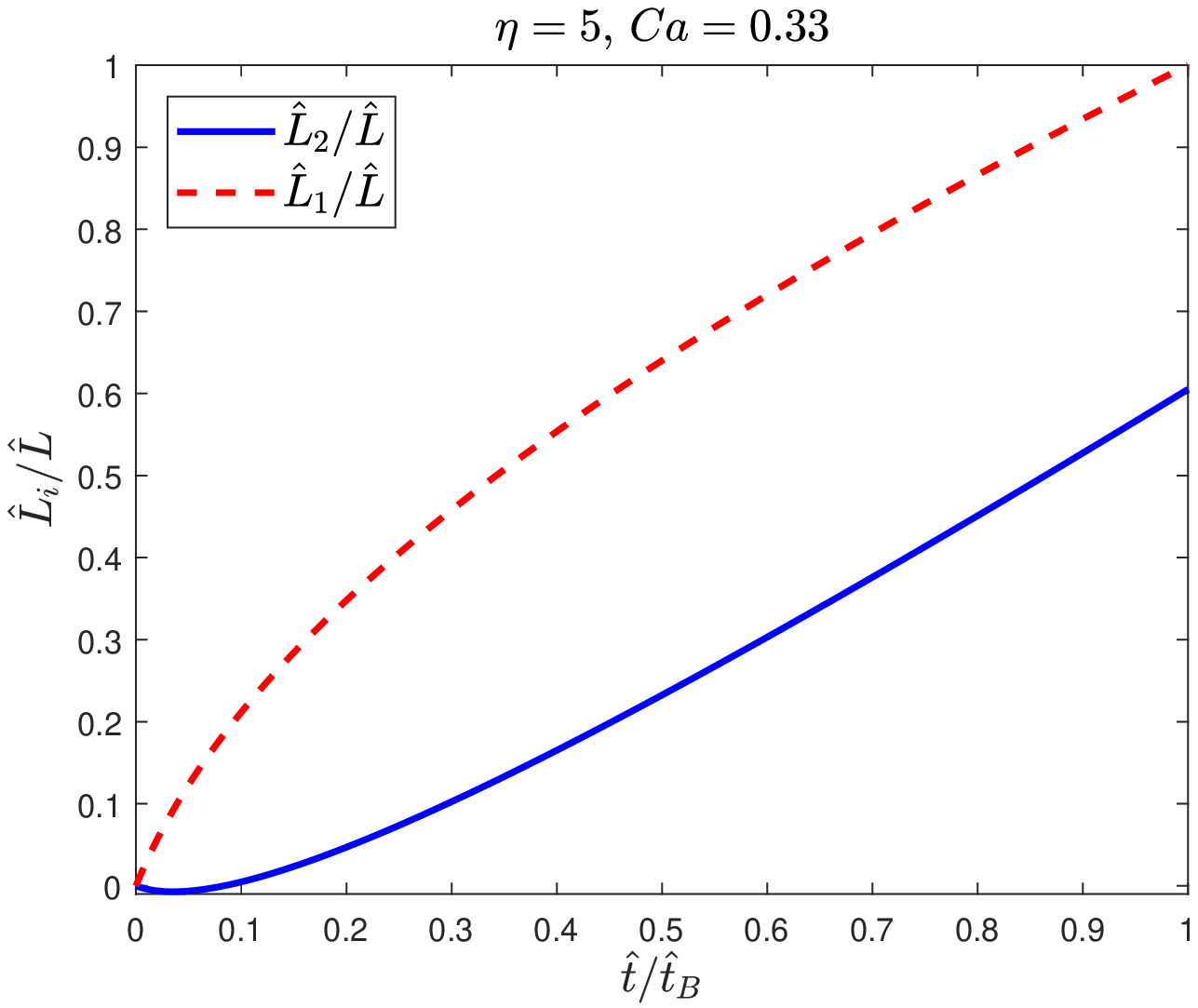}
		\end{minipage}
        \label{k_2_g}
	}
	\subfigure[]{
		\begin{minipage}[t]{0.3\linewidth}
			\includegraphics[width=1.8in]{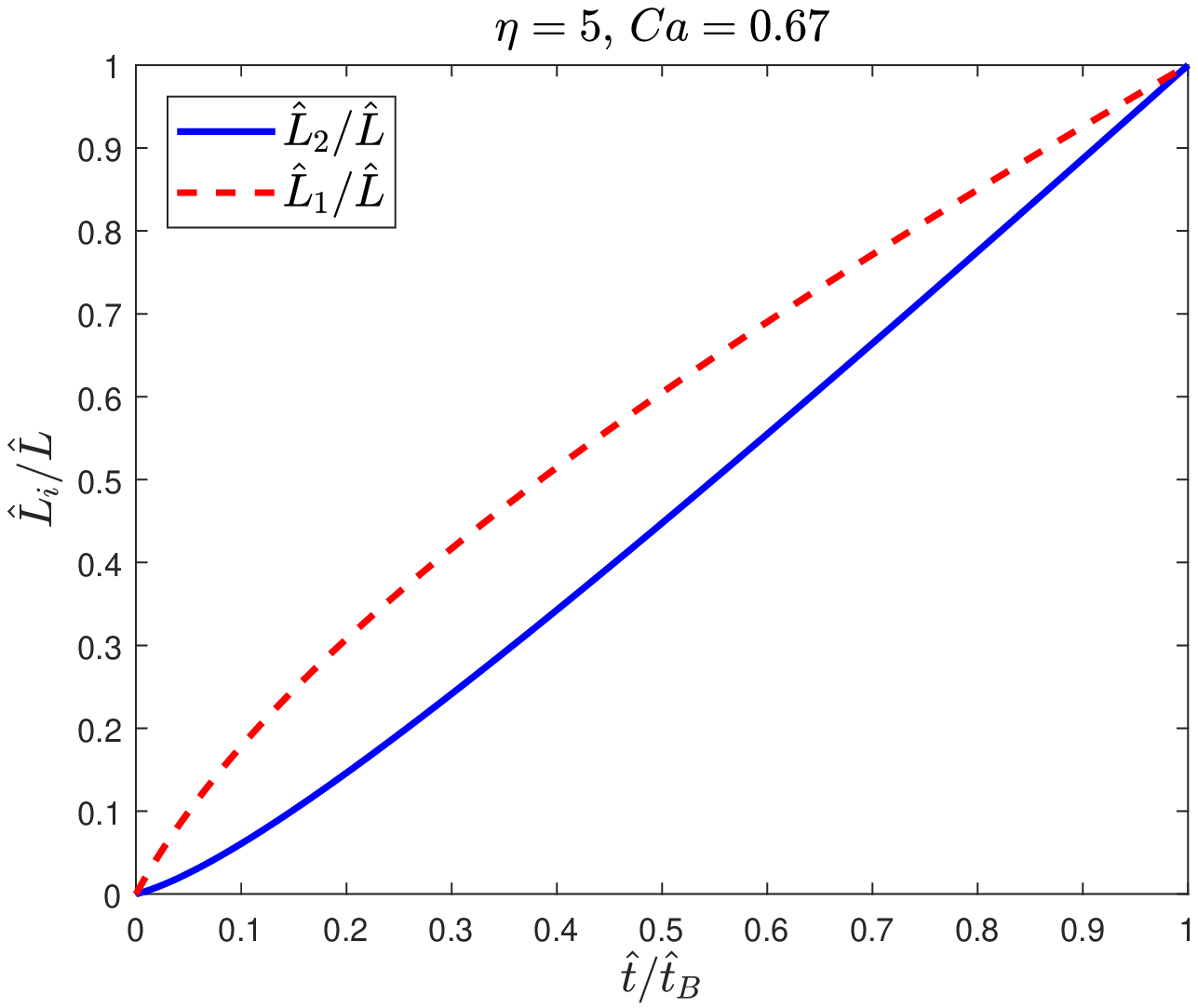}
		\end{minipage}
        \label{k_2_h}
	}
	\subfigure[]{
		\begin{minipage}[t]{0.3\linewidth}
			\centering
			\includegraphics[width=1.8in]{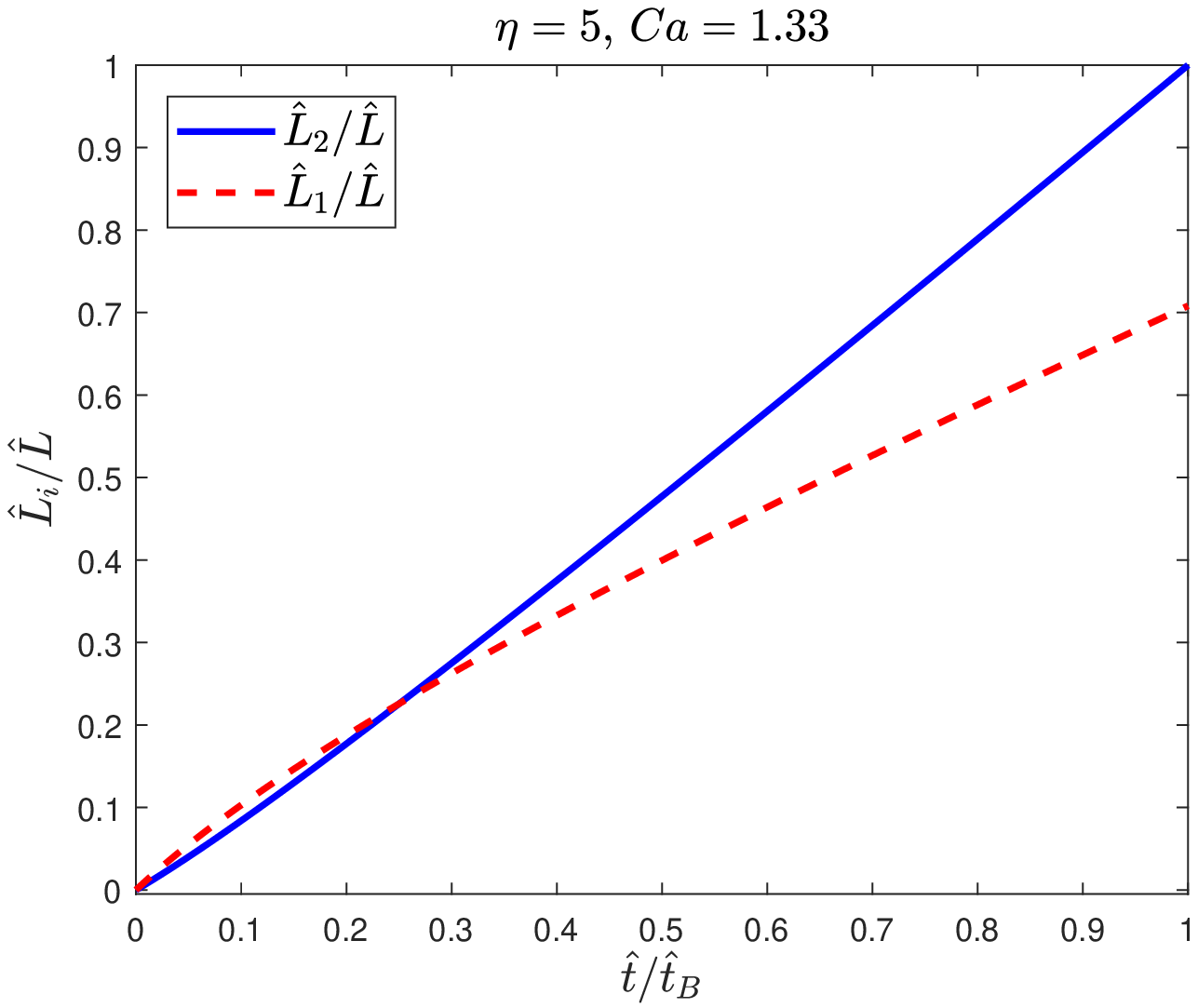}
		\end{minipage}
        \label{k_2_i}
	}
	\centering
	\caption{The penetration lengths of wetting fluids in the wide and narrow channels with the radius ratio $\hat{r} = 2$.}
	\label{Fig_A_N}
\end{figure}
\begin{figure}
	\centering
    \subfigure[]{
		\begin{minipage}[t]{0.4\linewidth}
			\centering
			\includegraphics[width=2.0in]{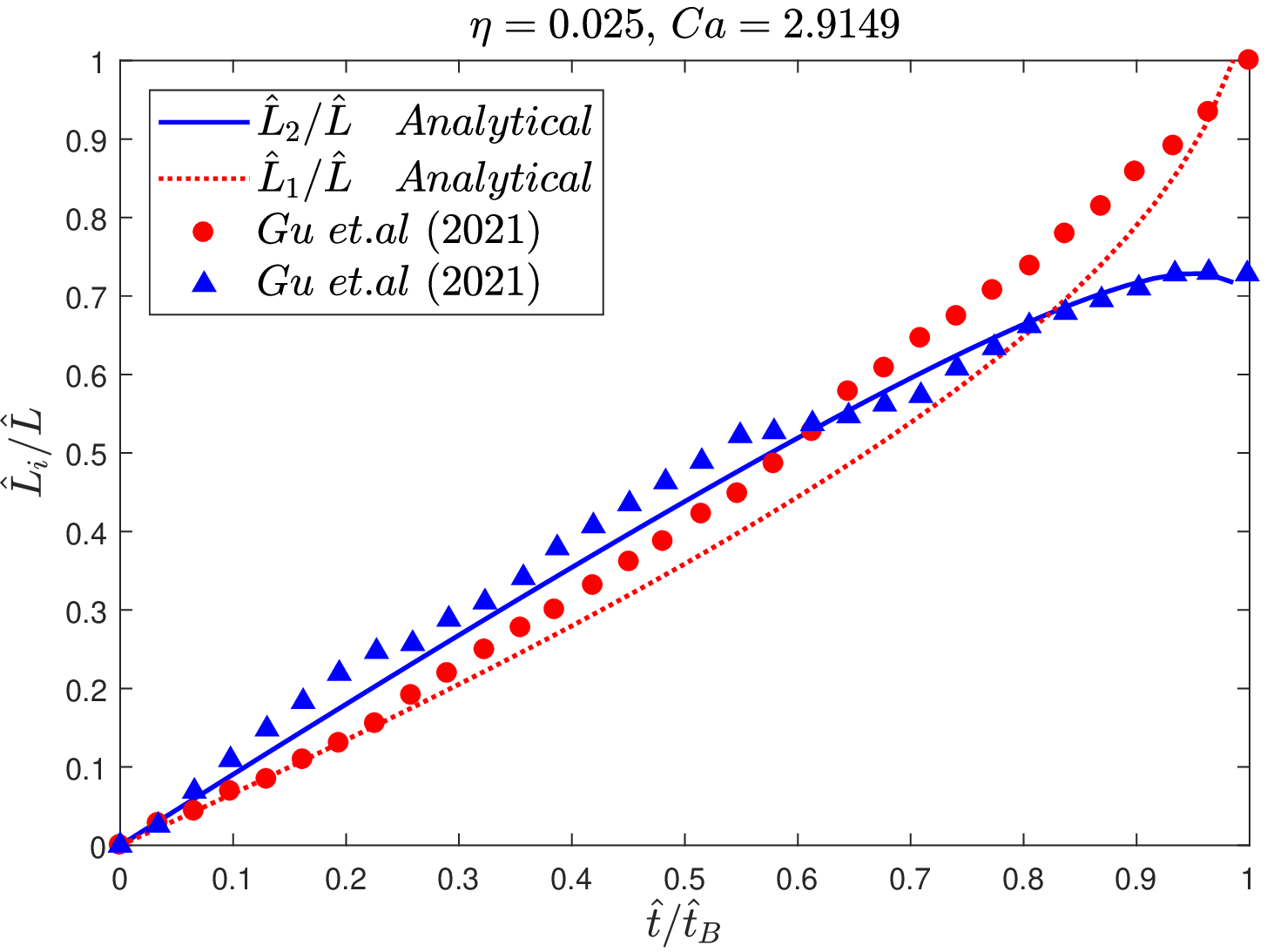}
		\end{minipage}
	}
    \subfigure[]{
		\begin{minipage}[t]{0.4\linewidth}
            \centering
			\includegraphics[width=2.0in]{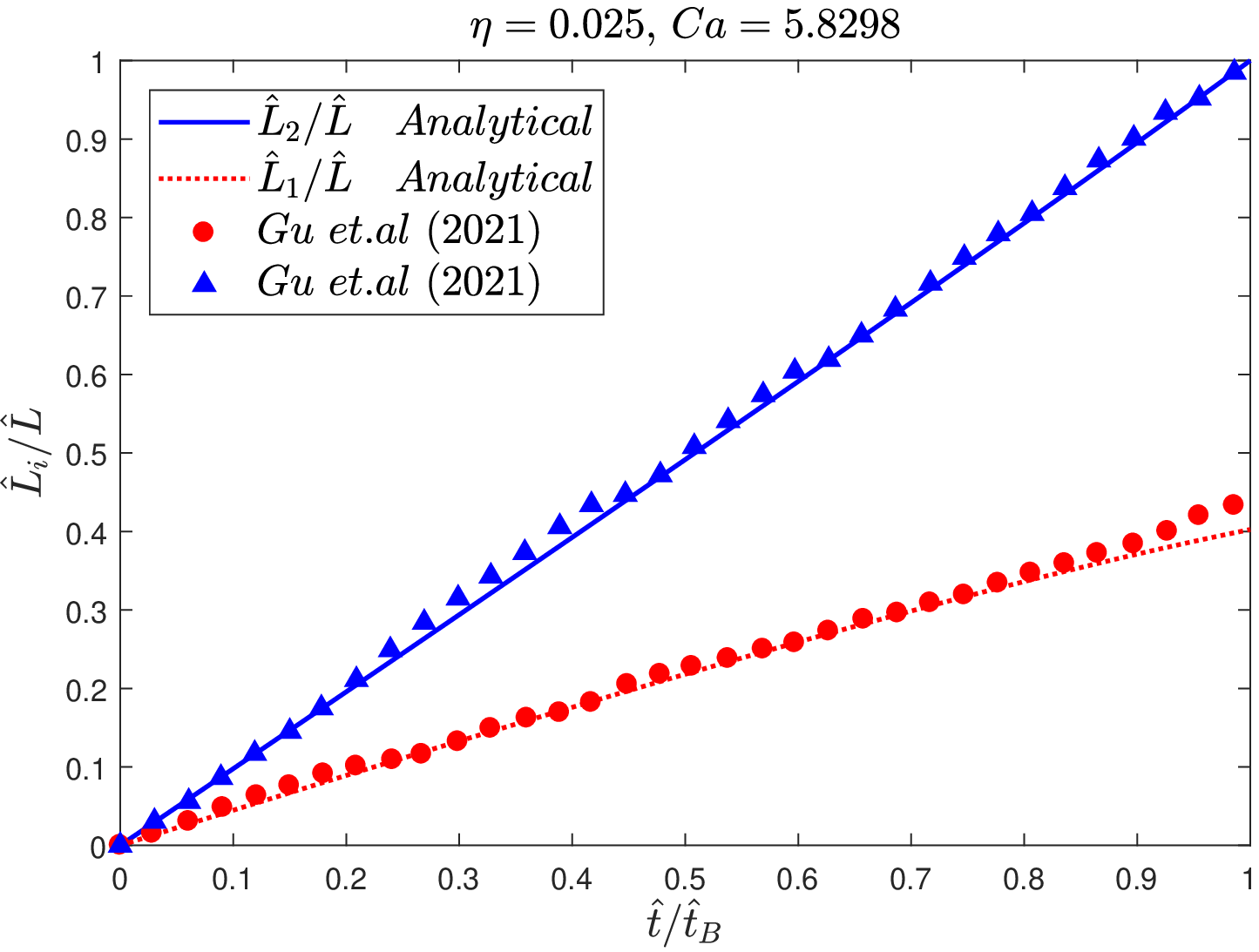}
		\end{minipage}
	}

    \subfigure[]{
		\begin{minipage}[t]{0.4\linewidth}
            \centering
			\includegraphics[width=2.0in]{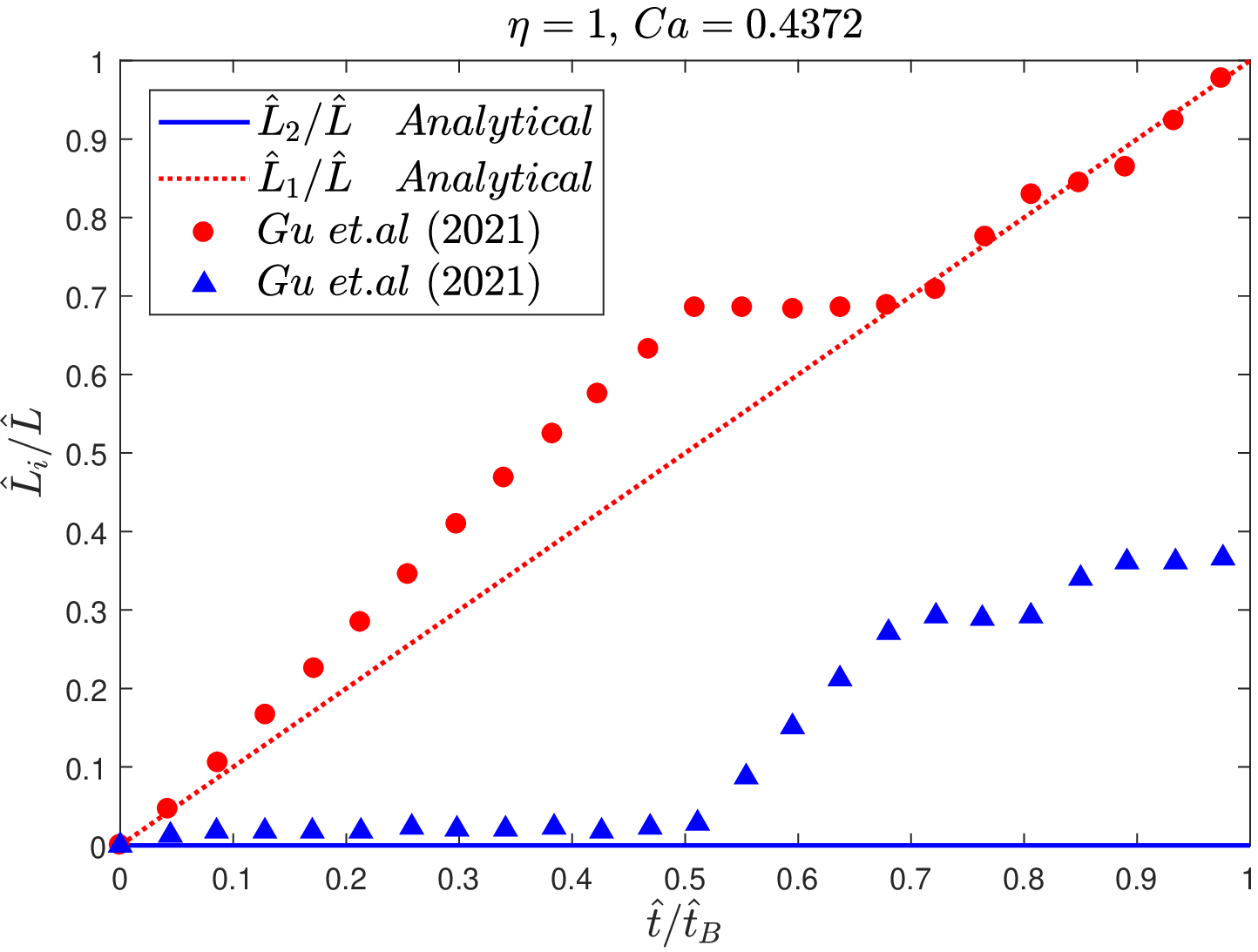}
		\end{minipage}
	}
	\subfigure[]{
		\begin{minipage}[t]{0.4\linewidth}
			\centering
			\includegraphics[width=2.0in]{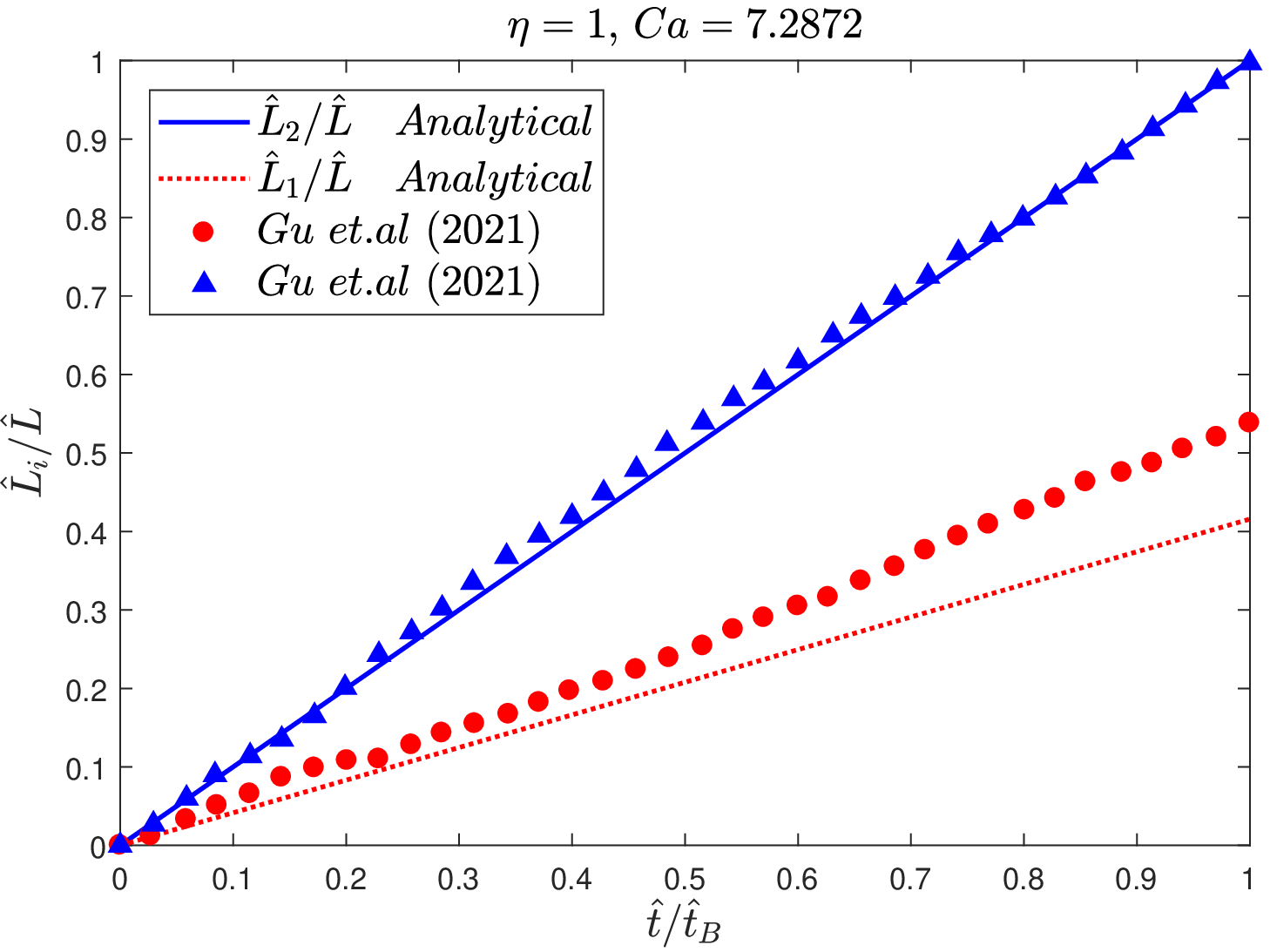}
		\end{minipage}
	}

	\subfigure[]{
		\begin{minipage}[t]{0.4\linewidth}
            \centering
			\includegraphics[width=2.0in]{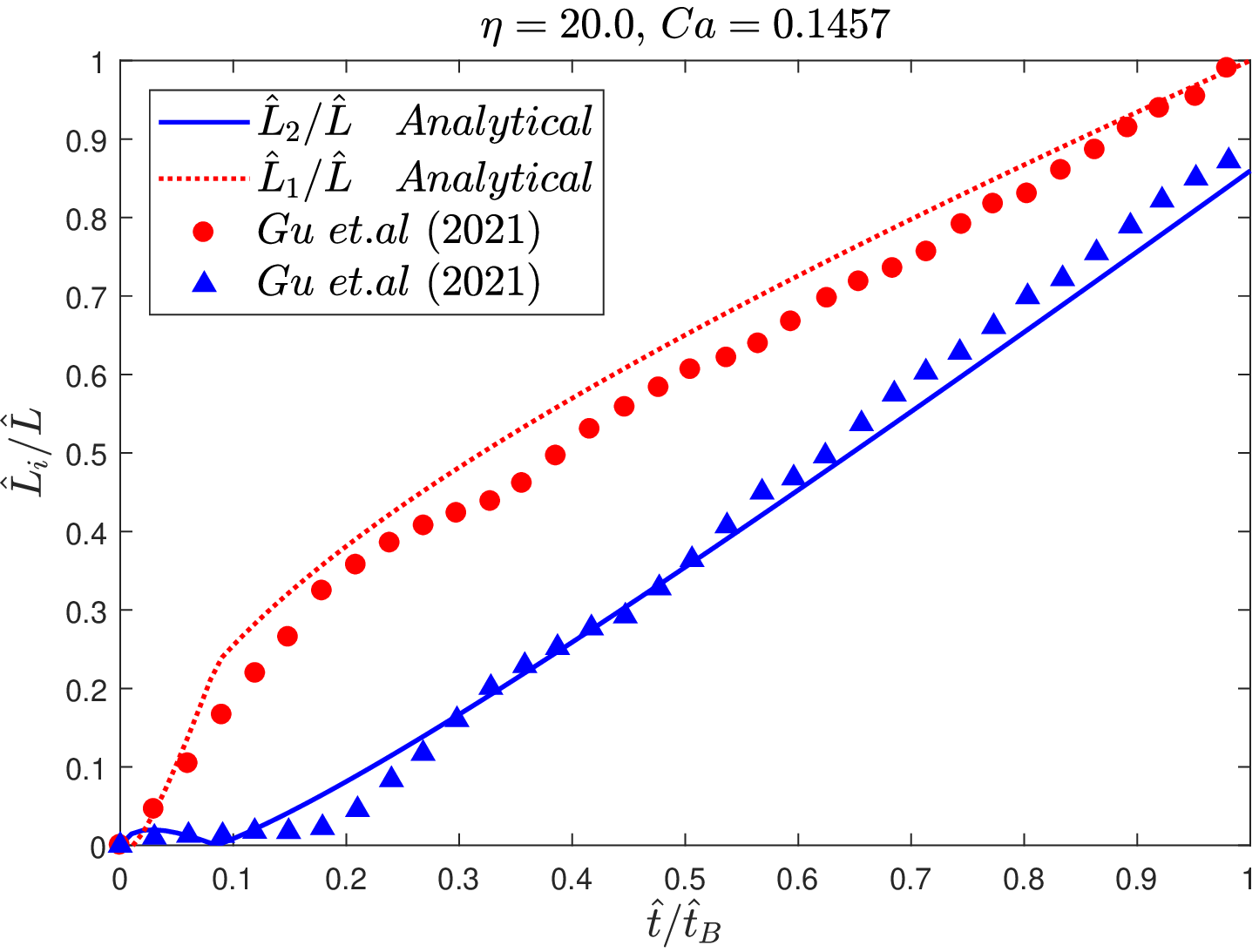}
		\end{minipage}
	}
\subfigure[]{
		\begin{minipage}[t]{0.4\linewidth}
			\centering
			\includegraphics[width=2.0in]{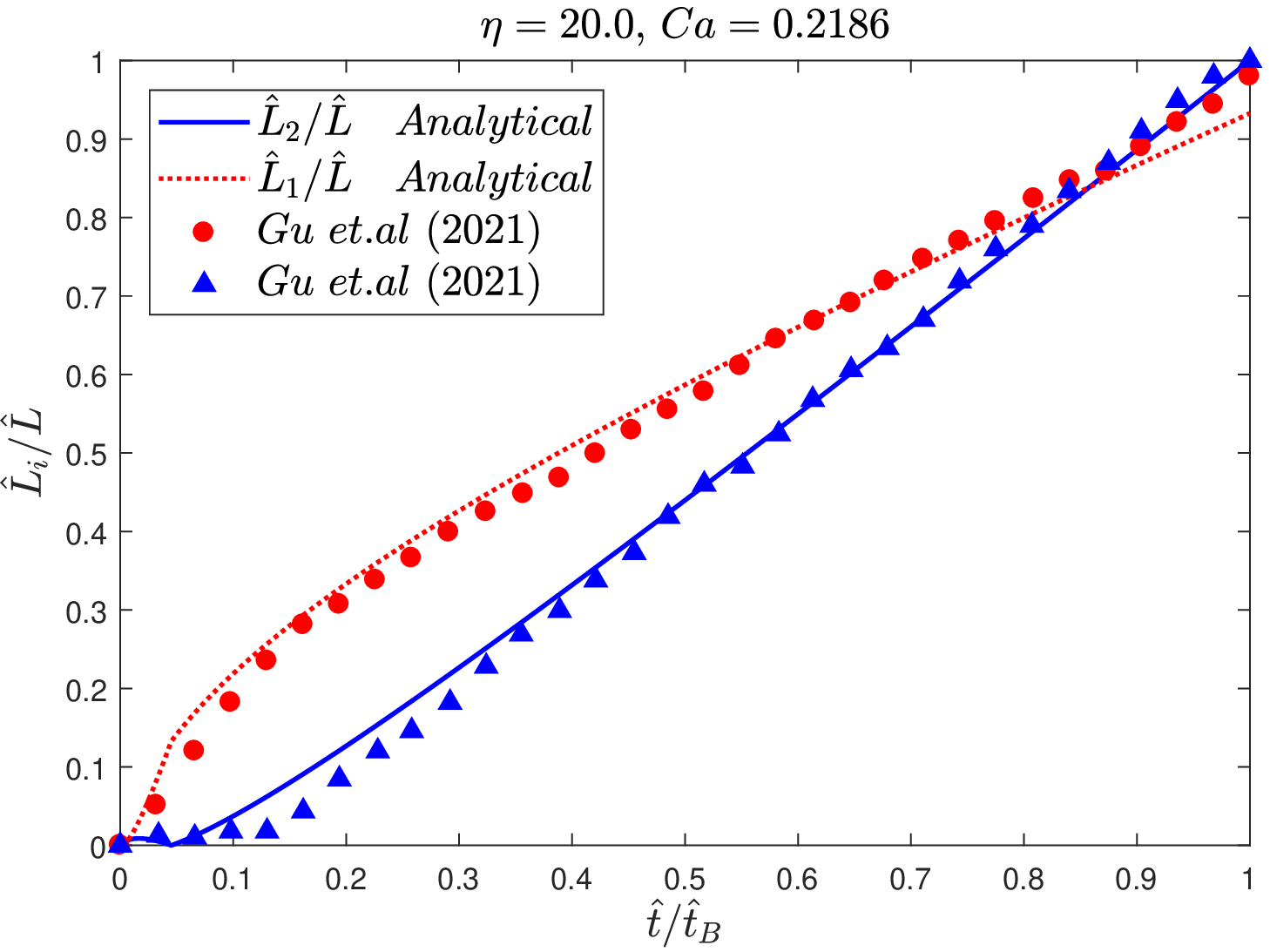}
		\end{minipage}
	}
	\centering
	\caption{The comparisons between analytical solutions (\ref{eq:L1_L2}) and numerical results of the penetration
length \citep{Gu2021} in the dual-permeability pore network at $\eta=0.025$ [(a) $Ca=2.9149$ and (b) $Ca=5.8298$],
$\eta=1$ [(c) $Ca=0.4372$ and (d) $Ca=7.2872$] and $\eta=20.0$ [(e) $Ca=0.1457$ and (f) $Ca=0.2186$]. }
	\label{FigL1}
\end{figure}

To further analyze the effects of the viscosity ratio and capillary number on the displacement process,
analytical solutions at several typical capillary numbers, viscosity ratios and radius ratios are
presented in Fig. \ref{k_2_a} where $\hat{r}=2$. From Fig. \ref{k_2_a}, one observes that for the
small viscosity ratio $\eta = 0.2$ and capillary number $Ca = 1.67$, the narrow channel with the
radius $r_{1}$ will be filled first. However, when $Ca$ increases to $6.67$, the wetting fluid
in the wide channel will first break through the channel, as shown in Fig. \ref{k_2_c}. It is also
found that at the critical capillary number $Ca_c = 3.33$, the menisci in the narrow and wide
channels will simultaneously reach the breakthrough position (point $B$), see Fig. \ref{k_2_b}.
In addition, for the cases with $\eta = 1$ and $\eta = 5$, one observes the similar phenomena from
Figs. \ref{k_2_d} - \ref{k_2_i}. These phenomena indicate that the displacement process mainly
depends on the relative relationship between viscous and capillary forces. For capillary dominated case,
the penetration length of wetting fluid in narrow channel is larger than that in wide channel during
breakthrough time since the capillary pressure is inversely proportional to the channel radius.
However, for the viscous dominated case, the viscous pressure drop in narrow channel is larger
than that in the wide channel, which results in a larger penetration length of wetting fluid in the wide channel.

Finally, the analytical results of pore-doublet model can also be applied to predict the displacement
in complex porous media. To this end, we select three typical viscosity ratio ratios, and present the
results of in Fig. \ref{FigL1}. From this figure, one finds that the analytical solution of penetration length $\hat{L}_{i}$ is close to the numerical data \citep{Gu2021}, especially for a large $Ca$. This also indicates that the preferential imbibition or preferential flow in porous media \citep{Gu2021,xie2021self,Fanli2022} can be predicted by the present analytical solution.
\section{Conclusions}\label{conclusion}
In this paper, an analytical study on the displacement of immiscible two-phase fluids in the pore doublet model is
conducted. We have first presented a mathematical model for the immiscible displacement in the
pore doublet model composed of two daughter channels with different radii, and then have obtained
the analytical solution of the mathematical model.

Based on the analytical solution, one can find that when $\hat{r} > 1$, the wetting fluid prefers to
invade the narrow channel for a small capillary number, and the breakthrough first occurs in the
narrow tube; while for a large capillary number, the wetting fluid will first break through the wide
channel. For the case with the critical capillary number, the menisci of wetting fluids in two branch
channels break through simultaneously. These different phenomena indicate that the competition between the viscous and capillary forces has a significant influence on the displacement process.

A phase diagram of critical capillary number $Ca_c$ in terms of viscosity ratio $\eta$ and radius ratio $\hat{r}$ is also presented to characterise the optimal displacement in the pore doublet. Moreover, we have also extended the two-dimensional results to three-dimensional case (see Appendices for details). Finally, we would also like to point out that the present study not only facilitates a fundamental understanding of the preferential penetration in the pore doublet system, but also provides a guideline to improve the efficiency of displacement in porous media.

\section*{Funding}This work was financially supported by the National Natural Science Foundation of China (Grants No. 12072127).
\section*{Declaration of interests}The authors declare that they have no known competing financial interests or personal relationships that could have appeared to influence the work reported in this paper.
\appendix\section{Mathematical model for immiscible displacement mechanisms in a three-dimensional pore doublet}\label{3D_model}
In this appendix, we focused on the displacement of immiscible two-phase fluids in the three-dimensional pore doublet composed of a pair of parallel daughter tubes which connects from entrance to the downstream. Similar to the analysis on the two-dimensional case, the pressure drop $\Delta p$ along the whole pore doublet can be expressed as
\begin{equation}
\Delta p=p_{A}-p_{B}=\Delta p_{i}+\Delta p_{c_{i}}=\frac{8q_{i}}{\pi r_{i}^4}\left[\mu_{w}L_{i}+\mu_{n}(L-L_{i})\right]-\frac{2\sigma \cos\theta}{r_{i}},
\label{eq:pressure_3D}
\end{equation}
where $\Delta p_{i}$ is the viscous pressure drop through tube $i$, and $\Delta p_{c_{i}}$ is the capillary pressure. Different from the two-dimensional pore doublet model, the volumetric flow rate in three-dimensional model is $q_{i}=\pi r_{i}^2\cdot u_{i}$, and the simultaneous breakthrough time is defined by $t_{B}=\pi(r_{1}^2+r_{2}^2)L/q$.
Introducing the following scaling parameters,
\begin{equation}
L_{c}=r_{1},\quad T_{c}=\frac{\pi r_{1}^2L}{q},\quad P_{c}=\frac{8\mu_{n}qL}{\pi r_{1}^4},\quad U_{c}=\frac{q}{\pi r_{1}^2},
\label{eq:3D_d}
\end{equation}
we rewrite the dimensionless form of pressure drop $\Delta p$ as
\begin{equation}
\begin{aligned}
\Delta \hat{p}=\hat{q}_{i}\left(\eta\frac{\hat{L}_{i}/\hat{L}}{\hat{r}_{i}^4}
+\frac{1-\hat{L}_{i}/\hat{L}}{\hat{r}_{i}^4}\right)-\frac{1}{Ca}\frac{1}{\hat{r}_{i}},i=1,2,
\end{aligned}
\label{eq:3D_dp}
\end{equation}
where $Ca=\frac{4q\mu_{n}L}{\pi\sigma\cos\theta L_{c}^3}$ is the capillary number, $\eta = \mu_{w}/\mu_{n}$ is the viscosity ratio. $\hat{q}_i$ is the dimensionless flow rate, and is defined by
\begin{equation}
\hat{q}_{i}=\hat{r}_{i}^2\frac{d(\hat{L}_{i}/\hat{L})}{d\hat{t}}.
\label{eq:qi}
\end{equation}
Substituting above equation into the mass conservation equation (\ref{eq:mass_conserve}), we have
\begin{equation}
\begin{aligned}
\hat{r}_{1}^2 \frac{d\left(\frac{\hat{L}_{1}}{\hat{L}}\right)}{d\hat{t}}+\hat{r}_{2}^2\frac{d\left(\frac{\hat{L}_{2}}{\hat{L}}\right)}{d\hat{t}}=1.
\end{aligned}
\label{eq:3D_mass}
\end{equation}

From (\ref{eq:3D_dp}) and (\ref{eq:3D_mass}), one can derive the mathematical model for the displacement in the three-dimensional pore doublet, which is composed of two coupled nonlinear ordinary differential equations,
\begin{subequations}
\begin{equation}
\frac{d\left(\frac{\hat{L}_{1}}{\hat{L}}\right)}{d\hat{t}}=\frac{\frac{1}{Ca}\left(1-\frac{1}{\hat{r}}\right)+\phi\left(\frac{\hat{L}_{2}}{\hat{L}}\right)}{\left[\phi\left(\frac{\hat{L}_{1}}{\hat{L}}\right)+\phi\left(\frac{\hat{L}_{2}}{\hat{L}}\right)\right]},
\end{equation}
\begin{equation}
\frac{d\left(\frac{\hat{L}_{2}}{\hat{L}}\right)}{d\hat{t}}=\frac{\frac{1}{Ca}\left(\frac{1}{\hat{r}}-1\right)+\phi\left(\frac{\hat{L}_{1}}{\hat{L}}\right)}{\hat{r}^2\left[\phi\left(\frac{\hat{L}_{1}}{\hat{L}}\right)+\phi\left(\frac{\hat{L}_{2}}{\hat{L}}\right)\right]},
\end{equation}
\label{eq:3D_ODE}
\end{subequations}
where $\phi\left(\frac{\hat{L}_{i}}{\hat{L}}\right)=\eta\frac{\hat{L}_{i}/\hat{L}}{\hat{r}_{i}^4}+\frac{1-\hat{L}_{i}/\hat{L}}{\hat{r}_{i}^4}$, $\hat{r}_1=1$, $\hat{r}_2=\hat{r}$.
\subsection{Analytical solution of mathematical model}\label{3D_solution}
Introducing the variables $x_{1}(\hat{t}) = \hat{L}_{1}/\hat{L}$, $x_{2}(\hat{t})=\hat{L}_{2}/\hat{L}$, $\bar{x}_{1}(\hat{t})=x_{1}(\hat{t})$ and $\bar{x}_{2}(\hat{t})=x_{2}(\hat{t})/\hat{r}^4$, one can rewrite (\ref{eq:3D_ODE}) as
\begin{subequations}
\begin{equation}
\bar{x}'_{1}(\hat{t})=\frac{Ca_{m}+\left(\eta-1\right)\bar{x}_{2}(\hat{t})+\frac{1}{\hat{r}^4}}{(\eta-1)\left[\bar{x}_{1}(\hat{t})+\bar{x}_{2}(\hat{t})\right]+1+\frac{1}{\hat{r}^4}},
\label{eq:3D_solve1}
\end{equation}
\begin{equation}
\hat{r}^6\bar{x_{2}}'(\hat{t})=\frac{-Ca_{m}+\left(\eta-1\right)\bar{x}_{1}(\hat{t})+1}{(\eta-1)\left[\bar{x}_{1}(\hat{t})+\bar{x}_{2}(\hat{t})\right]+1+\frac{1}{\hat{r}^4}},
\label{eq:3D_solve2}
\end{equation}
\label{eq:3D_solve}
\end{subequations}
where $Ca_{m}=\frac{1}{Ca}\left(1-\frac{1}{\hat{r}}\right)$. Combining (\ref{eq:3D_solve1}) and (\ref{eq:3D_solve2}), the equation of $\bar{x}_{2}(\hat{t})$ can be derived,
\begin{equation}
\frac{\eta-1}{2}\hat{r}^6(\hat{r}^6-1)\bar{x}_{2}^2(\hat{t})-\left[(\eta-1)\hat{t}\hat{r}^6+\hat{r}^6+\hat{r}^2\right]\bar{x}_{2}(\hat{t})+\frac{\eta-1}{2}\hat{t}^2+(1-Ca_{m})\hat{t}=0.\\
\label{eq:3D_x2}
\end{equation}

Under the conditions of $\bar{x}_{1}(0)=0$ and $\bar{x}_{2}(0)=0$, one obtains the analytical solution of (\ref{eq:3D_solve2}),
\begin{equation}
\begin{cases}
\bar{x}_{1}(\hat{t})=\hat{t}-\hat{r}^6\bar{x}_{2}(\hat{t}), \\
\bar{x}_{2}(\hat{t})=\frac{-b-\sqrt{\Delta}}{2a},\\
\end{cases}
\label{eq:3D_solution}
\end{equation}
where
\begin{equation}
\begin{aligned}
&a=\frac{\eta-1}{2}\hat{r}^6(\hat{r}^6-1),\quad b=-\left[(\eta-1)\hat{t}\hat{r}^6+\hat{r}^6+\hat{r}^2\right],\quad c=\frac{\eta-1}{2}\hat{t}^2+(1-Ca_{m})\hat{t}, \\
&\Delta=b^2-4ac=(\eta-1)^2\hat{r}^6\hat{t}^2 + 2(\eta-1)\hat{r}^6(\hat{r}^2 + \hat{r}^6Ca_{m} + 1 - Ca_{m})\hat{t} + \hat{r}^4(\hat{r}^4+1)^2.\\
\end{aligned}
\label{eq:3D_Delta}
\end{equation}
From (\ref{eq:3D_solution}) one can determine, the penetration lengths of wetting fluids in two branches ($\eta\neq 1$),
\begin{equation}
\begin{cases}
\hat{L}_{1}=\hat{L}(\hat{t}-\hat{r}^6\frac{-b-\sqrt{\Delta}}{2a}), \\
\hat{L}_{2}=\hat{r}^4\hat{L}\frac{-b-\sqrt{\Delta}}{2a}.\\
\end{cases}
\label{eq:3D_L1_L2}
\end{equation}
We would also like to point out that when $\eta = 1$, the analytical solution is given by
\begin{equation}
\begin{cases}
\hat{L}_{1}=\hat{L}\frac{\hat{r}^4Ca_{m}+1}{\hat{r}^4+1}\hat{t}, \\
\hat{L}_{2}=\hat{L}\frac{\hat{r}^2(1-Ca_{m})}{\hat{r}^4+1}\hat{t},\\
\end{cases}
\label{eq:L1_L2_1}
\end{equation}
where $Ca > (1 - 1/\hat{r})$.
\section{Critical capillary number}\label{critical_Ca_3D}
Similar to the previous section \ref{critical_Ca}, we now consider the critical capillary number corresponding to the optimal displacement in three-dimensional pore doublet. Under the condition of $\hat{L}_1= \hat{L}_2 = \hat{L}$, we can derive the following critical condition from (\ref{eq:3D_L1_L2})
\begin{equation}
Ca_{c} = \frac{2\hat{r}(\hat{r}-1)}{(\eta + 1)(\hat{r}^2+1)},
\label{eq:Cam_3D3}
\end{equation}
and the critical time $\hat{t} = 1+\hat{r}^2$. Specifically, when $\eta = 1$, the critical capillary number is only related to the radius ratio of channels,
\begin{equation}
Ca_{c} = \frac{\hat{r}(\hat{r}-1)}{\hat{r}^2+1}.
\end{equation}
\bibliographystyle{jfm}
\bibliography{jfm-instructions}

\end{document}